%% file: 0Paper.tex
\documentclass[sigconf,nonacm]{acmart}

\usepackage{multirow}
\usepackage{booktabs}
\usepackage{eqparbox}
\usepackage{tabularx}
\usepackage{float}
\usepackage{algorithm}
\usepackage{algorithmicx}
\usepackage{algpseudocode}
\usepackage{amsmath}
\usepackage{enumitem}
\usepackage{array}
\usepackage[most]{tcolorbox}
\algnewcommand{\algorithmicinput}{\textbf{Input:}}
\algnewcommand{\Input}{\item[\algorithmicinput]}
\algnewcommand{\algorithmicoutput}{\textbf{Output:}}
\algnewcommand{\Output}{\item[\algorithmicoutput]}

\AtBeginDocument{%
  }

\begin{document}

\title{NOVA: A Verification-Aware Agent Harness for Architecture Evolution in Industrial Recommender Systems}

\author{\texorpdfstring{%
\renewcommand{\arraystretch}{0.88}
\begin{tabular}{c}
Shaohua Liu$^{*}$, Liang Fang$^{*}$, Yilong Sun$^{*}$, Shudong Huang$^{*\dagger}$, Qingsong Luo, Shaoxin Liu,\\
Xiaoyang Chen, Dongqiang Liu, Chuangang Ma, Zhenzhen Chai, Henghuan Wang, Shijie Quan,\\
Changyuan Cui, Zhangbin Zhu, Peng Chen, Wei Xu, Lei Xiao, Haijie Gu, Jie Jiang\\[1pt]
Tencent Inc.\\
\texttt{\{jimmyshliu,franklfang,elonsun,ericdhuang,qsongluo,orionsxliu,noreenchen,}\\
\texttt{dongcialiu,chuangangma,njzzchai,asherwang,justinquan,shoncui,defyzhu,}\\
\texttt{pengchen,davidxu,shawnxiao,jerrickgu,zeus\}@tencent.com}\\
[1pt]
\end{tabular}
\vspace{-10pt}
}{Shaohua Liu et al.}}
\renewcommand{\shortauthors}{Shaohua Liu et al.}

\hypersetup{
  pdfauthor={Shaohua Liu, Liang Fang, Yilong Sun, Shudong Huang, Qingsong Luo, Shaoxin Liu, Xiaoyang Chen, Dongqiang Liu, Chuangang Ma, Zhenzhen Chai, Henghuan Wang, Shijie Quan, Changyuan Cui, Zhangbin Zhu, Peng Chen, Wei Xu, Lei Xiao, Haijie Gu, Jie Jiang},
  pdfkeywords={recommender systems, architecture modification, semantic verification}
}

\begin{abstract}
Industrial advertising recommender systems are continually improved through architecture modifications, yet production iteration remains expert-intensive because coordinated changes to model topology, feature configuration, and interaction modules must satisfy strict interface, resource, and serving constraints. AutoML is limited to predefined search spaces, while generic coding agents verify runnability rather than recommender-specific semantic validity. Executable candidates may therefore violate architectural contracts, while the lack of structured reuse of semantic diagnostics and evaluation outcomes can lead to repeated invalid or ineffective modifications.

We present \textsc{NOVA}, a verification-aware agent harness that organizes production architecture modification as multi-round search over concrete implementations within a fixed evaluation budget. At each round, NOVA generates multiple candidates under production constraints, rejects semantic violations, and ranks the valid survivors for local testing and offline evaluation. Across rounds, trajectory memory synthesizes semantic diagnostics, local-test outcomes, and offline metric changes into modification directions and forbidden patterns that guide subsequent search. Under the same maximum offline-evaluation budget for automated methods, NOVA achieves the highest effective pass rate, reaching $53.3\%$ on ScaleUp and $51.7\%$ on Literature-to-Production tasks. In a production A/B test covering $5\%$ of traffic in an advertising system serving over one billion users, the selected Literature-to-Production candidate yields GMV gains of $+1.25\%$, $+1.70\%$, and $+2.02\%$ across three major pCVR objectives, with corresponding relative reductions in absolute pCVR bias of $58.8\%$, $66.7\%$, and $37.3\%$, respectively.
\end{abstract}



\maketitle
\begingroup
\renewcommand{\thefootnote}{\fnsymbol{footnote}}
\footnotetext[1]{These authors contributed equally to this work.}
\footnotetext[2]{Corresponding author.}
\endgroup

\input{1Introduction.tex}
\input{2RelatedWork.tex}
\input{3ProblemDefinition.tex}
\input{4Methodology.tex}
\input{5Experiments.tex}
\input{6Conclusion.tex}

\bibliographystyle{ACM-Reference-Format}
\bibliography{7References}

\clearpage
\appendix
\input{8Appendix.tex}

\end{document}

%% file: 1Introduction.tex
\section{Introduction} \label{introduction}
Large-scale recommender systems have advanced through architectures that progressively enrich feature and behavior modeling. FM and FFM~\cite{rendle2010factorization,juan2016field} model explicit feature interactions; Wide \& Deep, DeepFM, and DCN~\cite{cheng2016wide,guo2017deepfm,wang2017dcn} learn higher-order interaction patterns; and DIN, DIEN, and SIM~\cite{zhou2018din,zhou2019dien,pi2020sim} incorporate users' behavioral histories. Recent production-oriented backbones, including RankMixer, TokenMixer-Large, MixFormer, HyFormer, and OneTrans~\cite{zhu2025rankmixer,jiang2026tokenmixer,huang2026mixformer,huang2026hyformer,zhang2026onetrans}, further improve recommendation quality by scaling model capacity and jointly modeling feature interactions and behavior sequences.

Despite these advances, architecture evolution in production remains slow, costly, and highly dependent on expert judgment. Engineers must first interpret recent research and translate a promising architectural idea into coordinated changes to the model topology, feature configuration, and interaction modules, while preserving compatibility with existing training and serving interfaces. They must then perform local testing, run offline training, analyze delayed results, and decide whether the candidate should proceed to online testing. Failed candidates consume substantial engineering time and GPU resources, while online A/B testing is constrained by business risk. Consequently, production teams can evaluate only a small fraction of plausible architecture modifications, making the overall process difficult to scale.

Scaling production architecture evolution therefore requires more effective use of each trial. Invalid candidates should be rejected before costly training, while evidence from attempted modifications should be retained rather than discarded. Standard local testing can expose compilation and shape errors, but some candidates remain runnable while violating recommender-specific architecture contracts. We term these \emph{semantic silent failures}; examples include mismatched feature-token mappings, missing sequence masks, incorrect logit-fusion paths and target leakage. They differ from \emph{effectiveness failures}, which preserve semantic validity and executability but fail to improve offline AUC. Semantic verification can reject the former before expensive training, whereas the latter can be identified only through offline evaluation.

Existing automation does not fully address this setting. AutoML, HPO, and conventional NAS search scalar hyper-parameters or predefined operator spaces, whereas production upgrades often require coordinated cross-module modifications under hard interface and resource constraints. Generic coding agents use execution feedback to repair implementations, while recent agentic systems reuse evaluation outcomes to refine subsequent solutions. However, they typically do not verify recommender-specific semantics before training or distinguish among semantic errors, local-test failures, and ineffective offline results when guiding later rounds (Section~\ref{sec:related}).

To address these challenges, we present \textsc{NOVA}, a verification-aware agent harness that formulates production architecture modification as multi-round candidate search under a fixed evaluation budget and production constraints. Unlike experiment loops driven primarily by execution and metric feedback, NOVA makes recommender-specific semantic validity an explicit decision stage and couples it with within-task trajectory feedback. At each round, NOVA generates multiple candidate modifications, rejects candidates that violate recommender-specific architecture semantics, and ranks the valid survivors for local testing and offline evaluation. The resulting review diagnostics, local-test outcomes, and offline metric changes are recorded in trajectory memory and synthesized into weak-component attributions, modification directions, and forbidden patterns that guide subsequent proposal generation, while task-specific and historical evidence support candidate ranking. At the start of each task, NOVA loads a knowledge-base snapshot that supplies fixed prior evidence throughout the search.

In summary, this paper makes the following contributions.

\begin{itemize}[leftmargin=*]
\item \textbf{A production-constrained architecture modification framework.}
We introduce \textsc{NOVA}, a verification-aware agent harness that organizes production architecture modification as multi-round, multi-candidate search over concrete implementations spanning computation graphs, feature configurations, and structural parameters. Candidates are generated subject to production constraints and evaluated in separate stages for semantic validity, executability, and offline effectiveness.

\item \textbf{Quality Review and within-task trajectory-guided search.}
We couple Quality Review with Trajectory Memory and Structured Feedback to form a verification-aware search loop. Quality Review rejects implementations that violate recommender architecture contracts before local testing, then ranks the semantic survivors using task-local trajectory history and a fixed historical knowledge base. Trajectory Memory records semantic diagnostics, local-test outcomes, and offline metric changes and synthesizes them into weak-component attributions, modification directions, and forbidden patterns that guide subsequent candidate generation.

\item \textbf{Industrial evaluation from offline search to online deployment.}
We evaluate NOVA against AutoML and coding-agent baselines on two production tasks under the same maximum offline-evaluation budget and include a retrospective human reference. NOVA achieves the highest effective pass rate on both tasks, reaching $53.3\%$ on ScaleUp and $51.7\%$ on Literature-to-Production. In a $5\%$-traffic production A/B test, the selected Literature-to-Production candidate yields GMV gains of $1.25$--$2.02\%$ and relative reductions in absolute pCVR bias of $37.3$--$66.7\%$ across three major objectives in an advertising system serving over one billion users.
\end{itemize}

%% file: 2RelatedWork.tex
\section{Related Work}
\label{sec:related}

NOVA draws on three lines of research: agentic recommender optimization, AutoML and architecture search, and coding or automated R\&D agents.

\paragraph{Agentic recommender optimization.}
Recent systems show that LLM agents can participate in production recommender optimization. YouTube's \textit{Self-Evolving Recommendation System}~\cite{wang2026self}, AgenticRecTune~\cite{wu2026agenticrectune}, and Meta's Ranking Engineer Agent (REA)~\cite{kumar2026rea} support proposal generation, configuration tuning, and ads-ranking experimentation. NOVA focuses on architecture modification in production recommenders. It verifies semantics before offline training and draws on two evidence sources: a fixed historical knowledge base and within-task trajectory memory. After each round, trajectory memory records semantic diagnostics, local-test results, and offline metric changes to guide candidate generation, verification, and ranking in later rounds.

\paragraph{AutoML, NAS, and HPO.}
AutoML, neural architecture search, and hyper-parameter optimization efficiently explore explicit spaces of hyper-parameters and architectures~\cite{thornton2013autoweka,feurer2015autosklearn,akiba2019optuna,pham2018enas,liu2019darts,real2019regularized}. Although these methods can search over predefined topology choices, they are less suited to coordinated changes across a production codebase, such as jointly updating feature schemas, interaction modules, training objectives, and serving interfaces. NOVA instead searches over executable cross-module modification plans while enforcing production constraints.

\paragraph{Coding and automated R\&D agents.}
Coding agents such as SWE-agent and OpenHands~\cite{yang2024sweagent,wang2025openhands}, together with reasoning and automated R\&D workflows such as ReAct and R\&D-Agent~\cite{yao2023react,yang2025rdagent}, combine code editing with iterative execution feedback. Such feedback supports debugging and task completion, but a runnable implementation may still violate domain-specific architectural semantics. NOVA adds multi-source semantic-rule verification and uses its reports, local-test results, and offline metric changes to guide later rounds away from invalid or ineffective modifications.

%% file: 3ProblemDefinition.tex
\section{Problem Formulation}
\label{sec:problem}

Given the initial architecture $A_0$ of a production recommender model, a user request $q$, and hard production constraints $\Omega$, the goal is to find an implemented architecture update that fulfills the request while satisfying $\Omega$. Offline selection uses only the objective specified in $q$.

\begin{figure*}[t]
    \centering
    \includegraphics[width=1\linewidth]{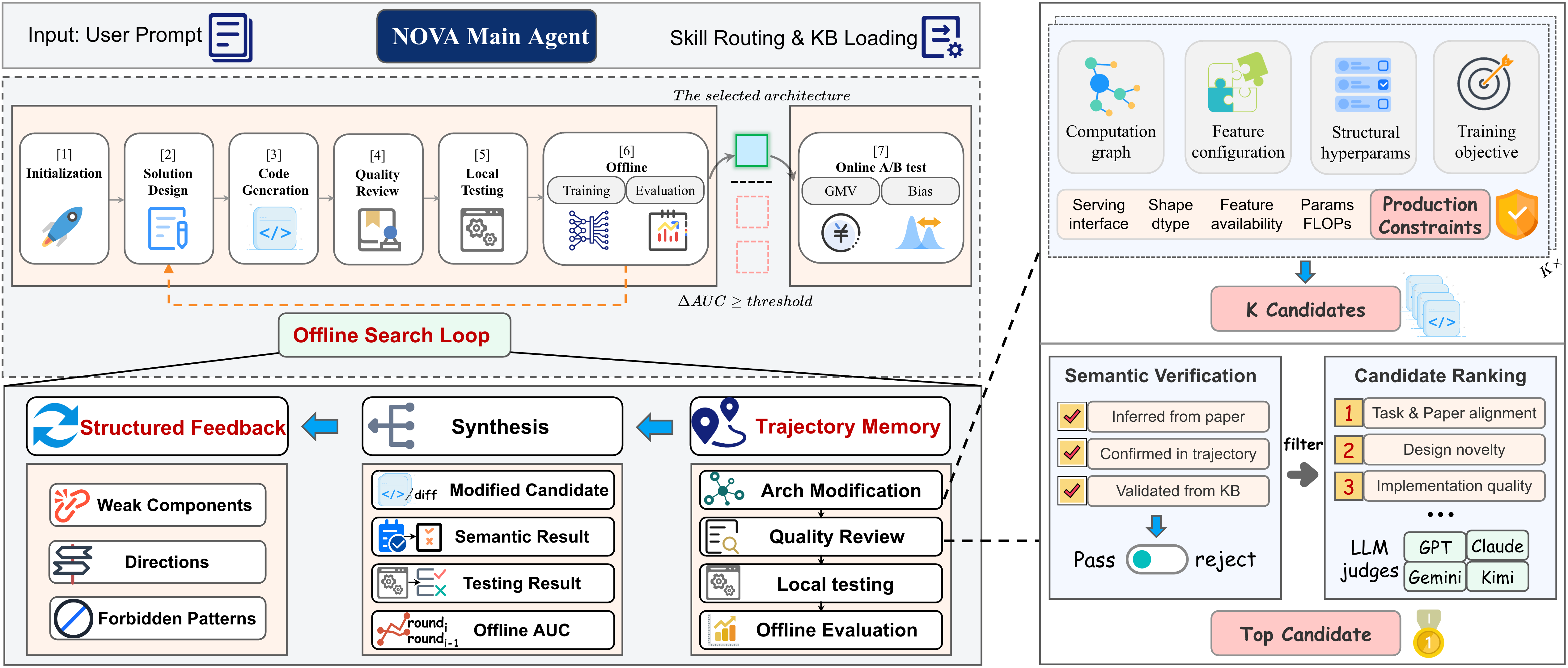}
    \caption{Overview of NOVA's verification-aware search loop. In each round, NOVA generates $K$ production-constrained candidate modifications, verifies their architecture semantics, ranks the valid candidates, and evaluates the selected candidate through local testing and offline training. Diagnostics and metric changes are recorded in trajectory memory and synthesized into structured guidance for the next round. The best candidate found offline proceeds to online validation.}
    \Description{System diagram of NOVA. A main agent coordinates initialization, solution design, code generation, quality review, local testing, offline evaluation, and online validation. Architecture modification generates production-constrained candidates; semantic verification and multi-LLM ranking select one candidate; trajectory memory records semantic, local-test, and offline outcomes and synthesizes weak components, modification directions, and forbidden patterns for later rounds.}
    \label{fig:nova_overview}
\end{figure*}

\paragraph{Architecture state and feasible candidates.}
At round $t$, the architecture state is
\begin{equation}
    A_t=(G_t,\phi_t,F_t).
\end{equation}
Here, $G_t$ is the computation graph, including interaction modules, prediction heads, and training-objective paths; $F_t$ is the feature configuration; and $\phi_t$ contains structural hyperparameters. Let $\mathcal E$ denote the modification-plan space. A plan $e_{t,j}\in\mathcal E$ may change any of these components. Applying $e_{t,j}$ to the current architecture $A_t$ yields
\begin{equation}
    \widetilde A_{t,j}=\mathrm{Apply}(A_t,e_{t,j}),
\end{equation}
where the candidate is retained only if it satisfies $\Omega$. These constraints cover tensor shapes and dtypes, feature availability, and parameter/FLOPs budgets. Candidates satisfying them are considered production-feasible and then undergo semantic verification and local testing for semantic validity and executability, respectively. Only candidates passing both stages proceed to offline effectiveness evaluation.

\paragraph{Feedback-guided architecture search.}
Each round generates multiple production-feasible candidates, rejects semantic violations, and ranks the survivors before evaluating one candidate. The resulting semantic diagnostics, local-test results, and offline metric changes are then reused to guide candidate generation, verification, and ranking in the next round. Under a fixed offline-evaluation budget, the search selects $A_{\mathrm{off}}^*$ as the production-feasible, semantically valid, and locally executable architecture with the best observed offline metric. It is then assessed in a production A/B test using GMV and prediction bias, which are validation metrics rather than search objectives.

%% file: 4Methodology.tex
\section{The NOVA Framework}
\label{sec:framework}

\subsection{Agent Harness Overview}
\label{subsec:overview}
Figure~\ref{fig:nova_overview} presents NOVA's end-to-end agent harness. The Main Agent selects a capability level and execution mode and coordinates seven stages from initialization through online evaluation. LLM agents perform architecture reasoning and review, while deterministic scripts execute testing, training, and metric collection. Each offline round connects three core modules: Architecture Modification generates production-constrained candidates, Quality Review filters and ranks them, and Trajectory Memory records their outcomes and synthesizes Structured Feedback for subsequent search.

\subsection{Architecture Modification}
\label{subsec:search}
At task entry, NOVA parses the user request $q$ and analyzes the production codebase to instantiate $A_0$. For Literature-to-Production tasks, Material Understanding further extracts the source paper's core architectural ideas and implementation details into the task context.

At round $t$, Solution Design uses the task request $q$, current architecture $A_t$, structured guidance $z_t$ from trajectory memory, and fixed prior $KB$ to produce $K$ modification plans $(e_{t,1},\ldots,e_{t,K})$ from $\mathcal{E}$. Code Generation applies each plan to $A_t$ under the production constraints $\Omega$, producing $\widetilde A_{t,j}$. The resulting candidate set is
\begin{equation}
    \mathcal C_t
    =
    \left\{
    c_{t,j}=(e_{t,j},\widetilde A_{t,j})
    \mid j=1,\ldots,K
    \right\}.
\end{equation}
Generating $K$ candidates provides alternative modification paths within each round, allowing the subsequent Quality Review stage to filter invalid implementations and select among the remaining candidates.

\subsection{Quality Review}
\label{subsec:quality-review}

Quality Review has two stages: semantic verification rejects candidates that violate applicable semantic rules, and evidence-guided ranking selects among the valid survivors.

\subsubsection{Semantic-Rule Verification}
At round $t$ of a task, NOVA assembles the active semantic-rule set $\mathcal L_t$ from paper requirements, semantic errors confirmed in earlier rounds (e.g., Table~\ref{tab:case1_trajectory}), and validated rules from $KB$. Each rule records its source, scope, and condition. For candidate $c_{t,j}$, $\mathcal L_{t,j}\subseteq\mathcal L_t$ denotes the applicable rules. NOVA performs semantic verification as
\begin{equation}
\label{eq:semantic-verification}
    r_{\mathrm{sem},t,j}
    =V_{\mathrm{sem}}(c_{t,j};\mathcal L_{t,j}).
\end{equation}
The report contains a decision in $\{\mathrm{pass},\mathrm{reject}\}$ and diagnostics that identify the violated rule and code location. Appendix~\ref{app:semantic-rules} gives representative rules from all three sources.

\subsubsection{Evidence-Guided Candidate Ranking}
Only semantically valid candidates enter ranking:
\begin{equation}
    \mathcal{S}_t
    =\left\{c_{t,j}\in\mathcal{C}_t
    \mid
    \ r_{\mathrm{sem},t,j}.\mathrm{decision}=\mathrm{pass}\right\}.
\end{equation}
Candidate ranking uses the latest production versions available during evaluation from four LLM families: GPT, Claude Opus, Gemini, and Kimi. Each judge independently scores a survivor on five dimensions: task alignment, paper alignment when applicable, consistency with effective evidence in $\mathcal O_t$ and $KB$, design novelty, and implementation quality. The judges are equally weighted, whereas the five dimensions use fixed, non-uniform weights held constant across tasks and rounds. Let $s_{t,j,d}^{(m)}$ be judge $m$'s score for candidate $c_{t,j}$ on dimension $d\in\mathcal{D}$, and let $w_d$ denote the corresponding dimension weight. The aggregate score is
\begin{equation}
    \label{eq:ranking-score}
    \rho_{t,j}=\frac{1}{4}\sum_{m=1}^{4}\sum_{d\in\mathcal{D}}
    w_d s_{t,j,d}^{(m)}.
\end{equation}
NOVA selects the highest-scoring survivor, denoted by $c_{t,j_t^*}=(e_{t,j_t^*},\widetilde A_{t,j_t^*})$, for local testing. It proceeds to offline evaluation only if the local tests pass.

\subsection{Trajectory Memory and Structured Feedback}
\label{subsec:memory-feedback}
Trajectory memory $\mathcal O_t$ contains the candidate records available at the start of round $t$. The selected candidate record is
\begin{equation}
    o_{t,j_t^*}=
    (c_{t,j_t^*},r_{\mathrm{sem},t,j_t^*},r_{\mathrm{loc},t},\delta J_t),
    \qquad
    \delta J_t=\widetilde J_t^*-J_t.
\end{equation}
Here, $J_t=J_{\mathrm{offline}}(A_t)$, $\widetilde J_t^*=J_{\mathrm{offline}}(\widetilde A_{t,j_t^*})$, and $\delta J_t$ is the gain over the current state. Each $o_{t,j}$ records one candidate in round $t$. Candidates rejected by semantic verification are stored as $(c_{t,j},r_{\mathrm{sem},t,j},\bot,\bot)$, while the selected candidate additionally records its local-test result and, after offline evaluation, $\delta J_t$. These records are appended to the cumulative history $\mathcal{O}_{t+1}$; semantically valid but unselected candidates are omitted. NOVA then synthesizes feedback for the next round:
\begin{equation}
    z_{t+1}
    =\mathrm{Synthesize}(\mathcal O_{t+1})
    =\left(
    z_{t+1}^{\mathrm{weak}},
    z_{t+1}^{\mathrm{dir}},
    z_{t+1}^{\mathrm{forbid}}
    \right).
\end{equation}
The three components identify weak components, recommend modification directions, and record ineffective modifications to avoid in subsequent rounds, respectively. During synthesis, an offline outcome inconsistent with the intended modification may trigger inspection of the selected candidate. A confirmed semantic error is converted into a semantic rule that remains available for verification in all subsequent rounds; the corresponding candidate and outcome remain in $\mathcal O_{t+1}$. Transient infrastructure failures are ignored.

\subsection{End-to-End NOVA Offline Search Loop}
\label{subsec:evolution-loop}

Algorithm~\ref{alg:nova} connects the three core modules without repeating their internal mechanisms. At each round, NOVA synthesizes the cumulative trajectory $\mathcal O_t$ into structured guidance $z_t$ and uses it to generate the candidate set $\mathcal{C}_t$. Quality Review then produces semantic reports, filters invalid candidates, and ranks the survivors. Local and offline outcomes are appended to $\mathcal{O}_{t+1}$. NOVA maintains both a working architecture $A_t$ for continued modification and the best observed architecture $A_{\mathrm{best}}$ for final selection. A semantic or local failure leaves the working architecture unchanged, whereas a candidate that reaches offline evaluation becomes the next working state. The search returns $A_{\mathrm{best}}$ when the improvement threshold is reached or the round budget is exhausted.

\begin{algorithm}[t]
\caption{End-to-End NOVA Offline Search Loop}
\label{alg:nova}
\footnotesize
\begin{algorithmic}[1]
\Require $q,A_0,\mathcal{E},\Omega,KB,R_{\max},K,\tau,J_{\mathrm{offline}}$, execution mode $\mu$
\Ensure Best feasible offline architecture $A_{\mathrm{off}}^*$ for online validation
\State $J_0 \gets J_{\mathrm{offline}}(A_0)$; \quad $A_{\mathrm{best}},J_{\mathrm{best}} \gets A_0,J_0$; \quad $\mathcal O_0\gets\emptyset$
\For{$t=0$ \textbf{to} $R_{\max}-1$}
  \State $z_t \gets \mathrm{Synthesize}(\mathcal O_t)$
  \State $(e_{t,1},\ldots,e_{t,K}) \gets \mathrm{SolutionDesign}(q,A_t,z_t,KB;\mathcal{E})$
  \State $\mathcal{C}_t\gets\{c_{t,j}=(e_{t,j},\mathrm{Apply}(A_t,e_{t,j};\Omega))\}_{j=1}^{K}$
  \State Compute $r_{\mathrm{sem},t,j}$ by Eq.~\eqref{eq:semantic-verification}, $\forall c_{t,j}\in\mathcal C_t$
  \State $\mathcal{S}_t \gets \{c_{t,j}\in\mathcal C_t\mid r_{\mathrm{sem},t,j}.\mathrm{decision}=\mathrm{pass}\}$
  \State $\mathcal{O}_{t+1}\gets\mathcal O_t\cup\{o_{t,j}=(c_{t,j},r_{\mathrm{sem},t,j},\bot,\bot)\mid r_{\mathrm{sem},t,j}.\mathrm{decision}=\mathrm{reject}\}$
  \If{$\mathcal{S}_t = \emptyset$}
    \State $A_{t+1},J_{t+1} \gets A_t,J_t$; \textbf{continue}
  \EndIf
  \State Compute $\rho_{t,j}$ by Eq.~\eqref{eq:ranking-score}, $\forall c_{t,j}\in\mathcal S_t$
  \State $c_{t,j_t^*}=(e_{t,j_t^*},\widetilde A_{t,j_t^*})
    \gets\operatorname*{arg\,max}_{c_{t,j}\in\mathcal S_t}\rho_{t,j}$
  \If{$\mu=\mathrm{Copilot}$}
    \State $\mathrm{HumanConfirm}(c_{t,j_t^*})$
  \EndIf
  \State $r_{\mathrm{loc},t} \gets V_{\mathrm{local}}(\widetilde{A}_{t,j_t^*})$
  \If{$r_{\mathrm{loc},t}.\mathrm{fail}$}
    \State $o_{t,j_t^*}\gets(c_{t,j_t^*},r_{\mathrm{sem},t,j_t^*},r_{\mathrm{loc},t},\bot)$
    \State $\mathcal{O}_{t+1}\gets\mathcal{O}_{t+1}\cup\{o_{t,j_t^*}\}$
    \State $A_{t+1},J_{t+1} \gets A_t,J_t$; \textbf{continue}
  \EndIf
  \State $\widetilde{J}_t^* \gets J_{\mathrm{offline}}(\widetilde{A}_{t,j_t^*})$
  \State $\delta J_t \gets \widetilde{J}_t^*-J_t$
  \State $o_{t,j_t^*}\gets(c_{t,j_t^*},r_{\mathrm{sem},t,j_t^*},r_{\mathrm{loc},t},\delta J_t)$
  \State $\mathcal{O}_{t+1}\gets\mathcal{O}_{t+1}\cup\{o_{t,j_t^*}\}$
  \State $A_{t+1},J_{t+1} \gets \widetilde{A}_{t,j_t^*},\widetilde{J}_t^*$
  \If{$\widetilde{J}_t^*>J_{\mathrm{best}}$}
    \State $A_{\mathrm{best}},J_{\mathrm{best}} \gets \widetilde{A}_{t,j_t^*},\widetilde{J}_t^*$
  \EndIf
  \If{$J_{\mathrm{best}}-J_0\ge\tau$} \textbf{break} \EndIf
\EndFor
\State \Return $A_{\mathrm{off}}^* \gets A_{\mathrm{best}}$
\end{algorithmic}
\end{algorithm}

\subsection{Agent Harness Implementation}
\label{subsec:implementation}

\paragraph{Task routing and control.}
The Main Agent routes modification plans, implementations, review reports, execution results, and structured feedback among stages while controlling $R_{\max}$, $K$, and $\tau$. NOVA assigns each task one of four capability levels: L1 for atomic structural tuning, L2 for constraint-aware ScaleUp, L3 for Literature-to-Production transfer, and L4 for open-ended architecture innovation. The level remains fixed within a task and determines its workflow and skills. Execution mode is selected separately according to validated skill coverage and operational risk: AutoRun proceeds autonomously, whereas Copilot requires human confirmation of the selected plan and implementation. Stages exchange only task-relevant structured artifacts.

\paragraph{Historical knowledge-base snapshot.}
At task entry, NOVA loads a task-specific snapshot $KB$ from specialized knowledge bases containing expert-validated effective modifications and failure patterns. The snapshot provides prior evidence for candidate generation, semantic verification, and candidate ranking, and remains unchanged during the current task. Reusable rules validated during the task may be committed afterward and appear only in future snapshots (Appendix~\ref{app:knowledge-base}).

\paragraph{Reasoning--execution separation.}
NOVA assigns architecture reasoning to LLM agents and operational tasks to deterministic scripts. LLM agents perform Material Understanding, Solution Design, Code Generation, Quality Review, and structured-feedback synthesis. Deterministic scripts retrieve materials, run local tests and training jobs, and collect metrics and logs. This separation keeps architecture reasoning flexible while making routine execution faster and deterministic. Only structured results, such as test status, metrics, and error summaries, are passed back to the LLM agents.

%% file: 5Experiments.tex
\section{Experiments}
\label{sec:experiments}
\subsection{Research Questions}
We evaluate NOVA around four research questions:

\begin{itemize}[leftmargin=*]
  \item \textbf{RQ1: Offline effectiveness.} Does NOVA achieve a higher rate of AUC-positive architecture modifications than automated baselines under the same maximum number of offline evaluations?

  \item \textbf{RQ2: Component contributions.} How does removing each NOVA component affect its effective pass rate and failure rates at different evaluation stages?

  \item \textbf{RQ3: Architecture transfer.} How does NOVA adapt and refine research architectures in production recommender code?

  \item \textbf{RQ4: Online impact.} Do the offline-selected modifications increase online GMV and reduce pCVR bias in production A/B tests?
\end{itemize}

\begin{figure}[t]
    \centering
    \includegraphics[width=\linewidth]{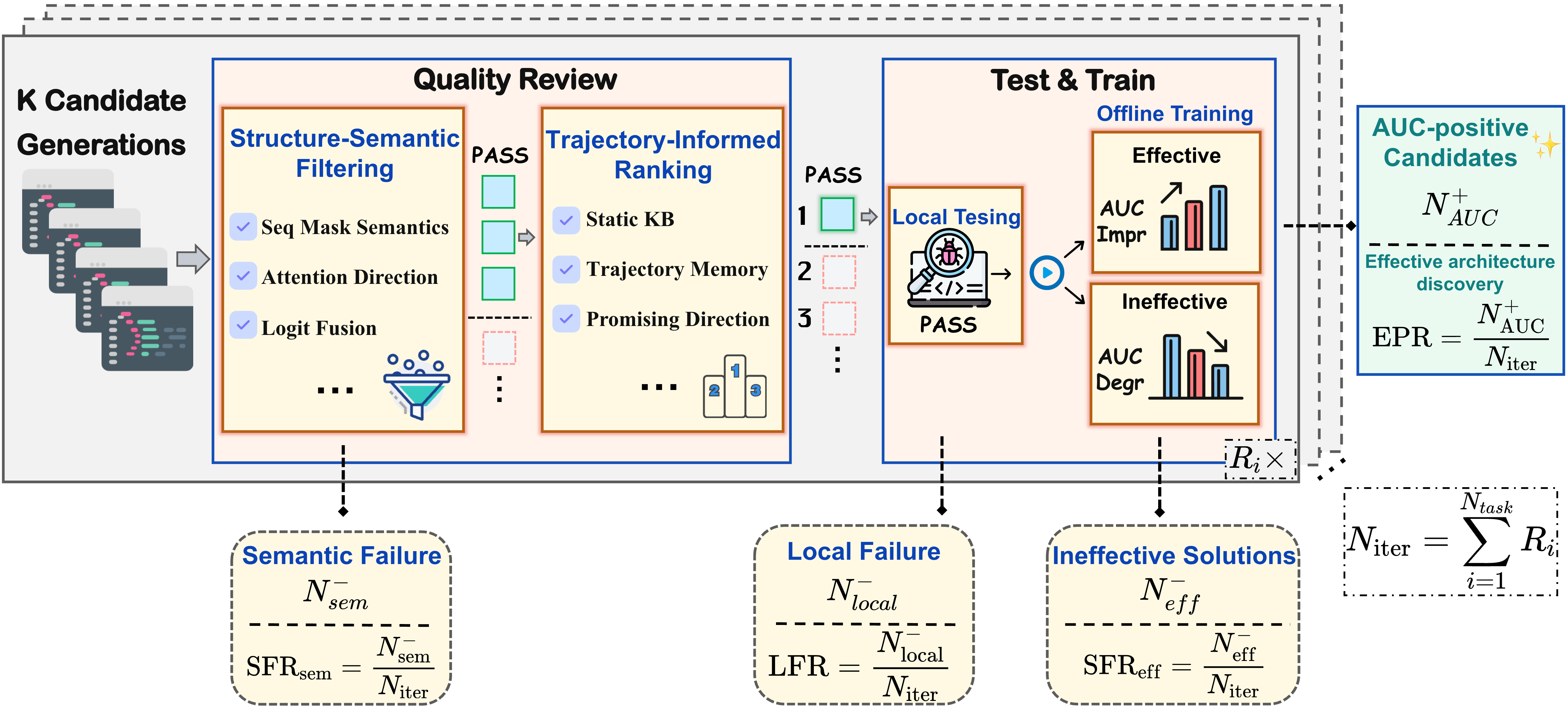}
    \caption{Task-level multi-round candidate funnel. Each task repeats the funnel for $R_i$ rounds. NOVA and each coding-agent baseline generate $K=4$ candidates in one candidate-generation process per round and evaluate at most one offline; all rates use $N_{\mathrm{iter}}=\sum_i R_i$ as their common denominator.}
    \Description{Candidate funnel for one task across multiple rounds. Each round generates four candidates, applies semantic verification and candidate ranking, evaluates at most one candidate through local testing and offline AUC, and records one of four outcomes: semantic failure, local-test failure, offline ineffectiveness, or an AUC-positive result.}
    \label{fig:candidate_funnel}
\end{figure}

\subsection{Experimental Settings}
\paragraph{Tasks.}
We evaluate NOVA on two architecture-modification settings. \textbf{L2 ScaleUp} uses a production RankMixer-style~\cite{zhu2025rankmixer} backbone and searches over coupled structural hyper-parameters, including \texttt{token\_cnt}, \texttt{token\_dim}, and the number of RankMixer layers, while keeping the total model size within a $\pm 10\%$ range of the production baseline. \textbf{L3 Literature-to-Production} evaluates whether a method can integrate literature-derived token-interaction modules, such as TokenMixer-Large~\cite{jiang2026tokenmixer} and MixFormer~\cite{huang2026mixformer}, into the production backbone under existing architectural and training-pipeline constraints.

\paragraph{Dataset.}
Experiments are conducted on a large-scale industrial advertising recommendation dataset collected from production traffic. The training corpus spans one month and contains billion-scale user--item interaction records. Each record is associated with over one thousand feature fields, covering both sequential and non-sequential signals. Every candidate model produced by every method is trained from scratch on the same dataset.

\paragraph{Budget and protocol.}
All LLM-dependent stages use \textbf{Claude Sonnet 4.6}~\cite{anthropic2026sonnet46}, except NOVA's four-model ranking ensemble described in Section~\ref{subsec:quality-review}. For each automated method, we run $N_{\mathrm{task}}=20$ tasks per setting. The L3 tasks are evenly split between TokenMixer-Large and MixFormer, with 10 tasks for each. Each task is capped at $R_{\max}=10$ rounds with the same early-stopping threshold $\tau$. Thus, $N_{\mathrm{iter}}=\sum_i R_i\leq200$, with small method-dependent differences due to early stopping. NOVA, its ablations, and coding-agent baselines generate $K=4$ candidates per round. For each coding-agent baseline, a fresh evaluator selects the Top-1 candidate from randomly shuffled candidates using only the task specification, production constraints, and candidate contents; NOVA uses its Quality Review, whereas Optuna-TPE follows its native sequential proposal mechanism. At most one candidate per round is evaluated offline, giving all automated methods the same maximum offline-evaluation budget. The human reference is retrospective and not budget matched.

\begin{table*}[t]
\caption{Main comparison on L2 ScaleUp and L3 Literature-to-Production tasks. EPR is the primary end-to-end metric; the three failure rates locate unsuccessful rounds in the evaluation funnel. Semantic failure is reported only for L3; ``--'' denotes an inapplicable method. The human row is retrospective rather than budget matched.}
\label{tab:main-l2-l3}
\small
\begin{tabularx}{\textwidth}{p{0.18\textwidth} *{8}{>{\centering\arraybackslash}X}}
\toprule
Method
& \multicolumn{3}{c}{L2 ScaleUp}
& \multicolumn{4}{c}{L3 Literature-to-Production} \\
\cmidrule(lr){2-4}\cmidrule(lr){5-8}
& LFR & $\mathrm{SFR}_{\mathrm{eff}}$ & EPR $\uparrow$
& $\mathrm{SFR}_{\mathrm{sem}}$ & LFR & $\mathrm{SFR}_{\mathrm{eff}}$ & EPR $\uparrow$ \\
\midrule
Human Reference
& 4.6\% & 46.2\% & 49.2\%
& 16.9\% & 43.7\% & 8.5\% & 31.0\% \\

OpenHands + Skills
& 66.7\% & 27.1\% & 6.3\%
& 45.7\% & 15.2\% & 21.7\% & 17.4\% \\

ReAct + Skills
& 62.5\% & 25.0\% & 12.5\%
& 39.6\% & 27.1\% & 25.0\% & 8.3\% \\

R\&D-Agent + Skills
& 37.5\% & 33.3\% & 29.2\%
& 19.6\% & 15.2\% & 37.0\% & 28.3\% \\

Optuna-TPE
& 82.8\% & 12.1\% & 5.2\%
& -- & -- & -- & -- \\

NOVA
& 1.7\% & 45.0\% & \textbf{53.3\%}
& 13.3\% & 11.7\% & 23.3\% & \textbf{51.7\%} \\
\bottomrule
\end{tabularx}
\end{table*}

\paragraph{Metrics.}
We report offline quality as baseline-relative AUC gain in percentage points:
\[
\Delta\mathrm{AUC}_{\mathrm{pp}}^{(0)}(A)
=100[\mathrm{AUC}(A)-\mathrm{AUC}(A_0)].
\]
An evaluated candidate is AUC-positive when $\Delta\mathrm{AUC}_{\mathrm{pp}}^{(0)}\geq0.10$ pp. A task terminates when the best observed baseline-relative gain reaches $\tau=0.15$ pp. We assess round-level reliability using four rates with $N_{\mathrm{iter}}$ as their common denominator:
\[
\begin{aligned}
\mathrm{SFR}_{\mathrm{sem}} &= \frac{N_{\mathrm{sem}}^-}{N_{\mathrm{iter}}}, &
\qquad \mathrm{LFR} &= \frac{N_{\mathrm{local}}^-}{N_{\mathrm{iter}}},\\
\mathrm{SFR}_{\mathrm{eff}} &= \frac{N_{\mathrm{eff}}^-}{N_{\mathrm{iter}}}, &
\mathrm{EPR} &= \frac{N_{\mathrm{AUC}}^+}{N_{\mathrm{iter}}}.
\end{aligned}
\]
Here, $N_{\mathrm{sem}}^-$ counts rounds with no semantically valid candidate; $N_{\mathrm{local}}^-$ counts rounds in which the selected candidate fails local testing; $N_{\mathrm{eff}}^-$ counts rounds that reach offline evaluation but fall below the AUC-positive threshold; and $N_{\mathrm{AUC}}^+$ counts rounds that produce an AUC-positive candidate. During search, NOVA's LLM-based $V_{\mathrm{sem}}$ filters candidates against the active semantic rules. For consistent evaluation across methods, all reported $\mathrm{SFR}_{\mathrm{sem}}$ values use post-hoc majority-vote labels from three expert engineers who independently audit every candidate (Appendix~\ref{app:semantic-rules}). For every automated L3 method, a round contributes to $N_{\mathrm{sem}}^-$ only when all $K=4$ candidates are labeled invalid. For the retrospective human reference, the same criterion applies to its single evaluated candidate. As illustrated in Figure~\ref{fig:candidate_funnel}, EPR is the primary end-to-end metric, while the three stage-specific failure rates partition unsuccessful rounds by the stage of failure.
\subsection{Baselines}

We compare NOVA with coding-agent and AutoML baselines and include a retrospective human reference. This is a system-level comparison under a matched maximum offline-evaluation budget; LLM inference cost is not matched. All coding-agent baselines receive the same reusable skills for paper understanding, production-backbone analysis, and code editing.

\begin{itemize}[leftmargin=*,nosep]
    \item \textbf{Human reference.}
    We score auditable senior-engineer modification records using the same outcome definitions and AUC-positive threshold, but report them only as retrospective context because they were not produced under the matched search budget.

    \item \textbf{ReAct + Skills.}
    A ReAct~\cite{yao2023react} agent uses the shared skills and repeated execution feedback. It does not use NOVA's semantic verification or within-task trajectory feedback.

    \item \textbf{OpenHands + Skills.}
    An OpenHands~\cite{wang2025openhands} agent uses the shared skills and execution feedback. It does not use NOVA's semantic verification or within-task trajectory feedback.

    \item \textbf{R\&D-Agent + Skills.}
    We implement a budget-matched adaptation of R\&D-Agent~\cite{yang2025rdagent} that retains its Researcher--Developer organization and uses the shared skills. The Researcher proposes modifications from paper, code, and performance evidence; the Developer implements them and repairs execution failures. The adaptation does not include NOVA's semantic-rule verification or its within-task trajectory memory over semantic, local-test, and offline outcomes.

    \item \textbf{Optuna-TPE.}
    Optuna-TPE~\cite{bergstra2011tpe} searches learning rate, hidden dimension, block number, dropout, token dimension, and token count under the same task-level stopping rule; it is applicable only to L2.
\end{itemize}

\subsection{Main Results for RQ1}

Table~\ref{tab:main-l2-l3} shows that NOVA improves EPR over R\&D-Agent, the strongest automated baseline, by $24.1$ percentage points on L2 ($53.3\%$ vs.\ $29.2\%$) and $23.4$ percentage points on L3 ($51.7\%$ vs.\ $28.3\%$); the human reference serves only as retrospective context.

\paragraph{L2 ScaleUp.}
The L2 ScaleUp task is conducted on a production RankMixer-style backbone. Although it appears to be simple parameter tuning, the target variables are structurally coupled; for example, \texttt{token\_dim}  must be divisible by \texttt{token\_cnt} due to the tokenization and interaction design~\cite{zhu2025rankmixer}. NOVA's $1.7\%$ LFR, compared with $37.5\%$--$82.8\%$ for the automated baselines, indicates the benefit of architecture-aware constraint handling over black-box search. Its EPR result shows that this reduction translates into more AUC-positive modifications rather than merely moving failures to offline evaluation.

\paragraph{L3 Literature-to-Production.}
L3 evaluates the integration of paper-derived modules into the production backbone under architectural and training-pipeline constraints. Against coding-agent baselines using the same skill bundle, NOVA achieves the highest EPR at $51.7\%$, showing that shared skills alone do not explain its system-level advantage. RQ2 further examines the contribution of within-task trajectory-guided iteration and other NOVA components.

\subsection{Ablation Study for RQ2}
\label{subsec:exp_ablation}
To answer RQ2, we ablate NOVA on the L3 setting under the same budget and evaluation protocol. Table~\ref{tab:ablation} groups the variants by architecture modification, quality review, within-task trajectory feedback, and the historical knowledge base.

\begin{table*}[t]
\centering
\small
\caption{Component-level ablation on the L3 task. EPR is the primary end-to-end metric; the failure rates show how unsuccessful rounds are distributed across evaluation stages. All $\mathrm{SFR}_{\mathrm{sem}}$ values use independent expert evaluation.}
\label{tab:ablation}
\begin{tabularx}{\textwidth}{p{0.22\textwidth} X c c c c}
\toprule
\textbf{Variant}
& \textbf{Ablation intervention}
& $\mathbf{SFR}_{\mathrm{sem}}$
& \textbf{LFR}
& $\mathbf{SFR}_{\mathrm{eff}}$
& \textbf{EPR} $\uparrow$ \\
\midrule

\textbf{NOVA (full)}
& Full system
& $13.3\%$ & $11.7\%$ & $23.3\%$ & $\mathbf{51.7\%}$ \\

\midrule
\multicolumn{6}{l}{\textit{Architecture Modification (Section~\ref{subsec:search})}} \\
\midrule
w/o Material Understanding
& Uses the raw paper without structured understanding artifacts
& $18.2\%$ & $13.6\%$ & $31.8\%$ & $36.4\%$ \\

w/o Solution Design
& Removes request interpretation, backbone analysis, and integration planning
& $36.8\%$ & $15.8\%$ & $21.1\%$ & $26.3\%$ \\

w/o Multi-Candidate Generation
& Sets $K=1$ instead of $K=4$
& $21.7\%$ & $30.4\%$ & $13.0\%$ & $34.8\%$ \\

\midrule
\multicolumn{6}{l}{\textit{Quality Review (Section~\ref{subsec:quality-review})}} \\
\midrule
w/o Semantic-Rule Verification
& Disables search-time $V_{\mathrm{sem}}$; all constraint-valid candidates enter ranking
& $24.2\%$ & $15.2\%$ & $39.4\%$ & $21.2\%$ \\

w/o Evidence-Guided Candidate Ranking
& Selects uniformly from $\mathcal{S}_t$ instead of using $\rho_{t,j}$
& $20.0\%$ & $17.1\%$ & $51.4\%$ & $11.4\%$ \\

\midrule
\multicolumn{6}{l}{\textit{Trajectory Memory and Structured Feedback (Section~\ref{subsec:memory-feedback})}} \\
\midrule
w/o Trajectory Memory and Structured Feedback
& Removes $\mathcal O_t$ and $z_t$ while retaining $KB$
& $42.9\%$ & $19.0\%$ & $14.3\%$ & $23.8\%$ \\

\midrule
\multicolumn{6}{l}{\textit{Historical Knowledge-Base Snapshot (Section~\ref{subsec:implementation})}} \\
\midrule
w/o Knowledge Base
& Sets $KB=\emptyset$ while retaining $\mathcal O_t$ and $z_t$
& $47.6\%$ & $33.3\%$ & $4.8\%$ & $14.3\%$ \\
\bottomrule
\end{tabularx}
\end{table*}

\paragraph{Architecture Modification.}
Removing Material Understanding, Solution Design, or Multi-Candidate Generation reduces EPR by $15.3$, $25.4$, and $16.9$ percentage points, respectively. The $K=1$ variant's round-level $\mathrm{SFR}_{\mathrm{sem}}$ is not directly comparable with the $K=4$ variants because semantic failure requires all candidates in a round to be invalid.

\paragraph{Quality Review.}
Removing Semantic-Rule Verification reduces EPR by $30.5$ percentage points; removing Evidence-Guided Candidate Ranking reduces it by $40.3$ points. In the ranking ablation, failures shift primarily to offline ineffectiveness ($\mathrm{SFR}_{\mathrm{eff}}=51.4\%$); in the verification ablation, semantic and effectiveness failures both increase.

\paragraph{Trajectory Memory, Structured Feedback, and Historical Knowledge Base.}
With the historical knowledge base retained, removing Trajectory Memory and Structured Feedback reduces EPR from $51.7\%$ to $23.8\%$, providing direct evidence that within-task trajectory-guided iteration is a major contributor beyond shared skills. Removing the historical knowledge-base snapshot while retaining $\mathcal O_t$ and $z_t$ reduces EPR by $37.4$ percentage points. In both variants, failures shift upstream: $\mathrm{SFR}_{\mathrm{sem}}$ rises to $42.9\%$ and $47.6\%$, while removing $KB$ also raises LFR to $33.3\%$.

\subsection{Production-Code Modification Case Study for RQ3}
\label{subsec:exp_case}
We report two L3 cases on a RankMixer-style backbone. Case 1 uses trajectory feedback to refine transferred TokenMixer-Large modules~\cite{jiang2026tokenmixer}. Case 2 extends MixFormer~\cite{huang2026mixformer} with gated multi-sequence fusion.

\paragraph{Case 1: Trajectory-guided correction of auxiliary loss.}
Objectives $o_1$--$o_3$ belong to the same pCVR model. The target is a baseline-relative AUC gain of at least $+0.15$ pp on $o_1$ without degrading $o_2/o_3$. Table~\ref{tab:case1_trajectory} traces six rounds.

\begin{table*}[t]
\centering
\scriptsize
\setlength{\tabcolsep}{3pt}
\renewcommand{\arraystretch}{1.03}
\caption{Six-round trajectory for Case 1. Values report baseline-relative $\Delta\mathrm{AUC}_{\mathrm{pp}}^{(0)}$ for the three pCVR objectives $o_1/o_2/o_3$. The target is $o_1\ge+0.15$ pp while keeping $o_2$ and $o_3$ non-negative.}
\label{tab:case1_trajectory}
\begin{tabularx}{\textwidth}{c p{0.30\textwidth} p{0.18\textwidth} X}
\toprule
\textbf{Round} & \textbf{Architecture modification} & \textbf{$\Delta$AUC (pp): $o_1/o_2/o_3$} & \textbf{Synthesized trajectory feedback} \\
\midrule
1 & Transfer Mixing-and-Reverting, PerToken-SwiGLU, and AuxLoss
  & $+0.09/+0.06/+0.05$ & \textbf{Weak:} $o_1$ remains below target; dense parameters exceed $2\times$ the backbone. \textbf{Direction:} isolate AuxLoss and revert parameter scaling before further changes. \\
2 & Remove AuxLoss
  & $+0.07/-0.03/-0.04$ & \textbf{Weak:} removing AuxLoss degrades $o_2/o_3$, while $o_1$ remains far below target. \textbf{Direction:} restore it at a lower weight and add $o_1$-specific supervision. \\
3 & Restore AuxLoss at $0.1$; add an $o_1$-specific auxiliary loss
  & $+0.08/+0.05/-0.06$ & \textbf{Weak:} the $o_1$-specific loss adds little gain, indicating a possible mismatch in its prediction/label indexing; the restored AuxLoss degrades $o_3$. \textbf{Direction:} inspect $o_1$ indexing and mask $o_3$ from AuxLoss. \\
4 & Fix $o_1$ auxiliary indexing; mask $o_3$ from AuxLoss
  & $+0.11/+0.02/-0.01$ & \textbf{Weak:} $o_1$ remains below target. \textbf{Direction:} add an $o_1$-specific token MLP and strengthen supervision. \textbf{Semantic rule:} code inspection confirms the $o_1$ indexing mismatch; each auxiliary loss must be routed to its design-specified objective. \\
5 & Add an $o_1$-specific token MLP; increase the $o_1$ AuxLoss weight
  & $-0.24/-0.06/-0.06$ & \textbf{Weak:} the MLP and higher loss weight degrade all objectives. \textbf{Forbid:} avoid repeating this joint MLP and loss-weight increase in the current task. \textbf{Direction:} revert to the Round 4 configuration and retune the $o_1$ AuxLoss weight. \\
6 & Revert to Round 4; retain corrected $o_1$ indexing and $o_3$ masking, and reduce the $o_1$ AuxLoss weight
  & $+0.15/+0.02/+0.09$ & \textbf{Outcome:} target reached without degrading $o_2/o_3$; stop. \\
\bottomrule
\end{tabularx}
\end{table*}

Although Round~3 passes the active rules and local testing, its limited $o_1$ gain and degradation on $o_3$ trigger code inspection. The inspection reveals that the $o_1$-specific auxiliary loss uses prediction and label indices from the wrong objective, an error absent from the active rule set. Round~4 corrects the indexing, masks $o_3$ from AuxLoss, and adds the objective-routing rule for later rounds. Round~5 then degrades all three objectives after jointly adding a token MLP and increasing the loss weight. Rather than continue from this result, Round~6 rolls back to Round~4, retains its indexing and masking corrections, and retunes only the $o_1$ AuxLoss weight to reach the target without degrading $o_2/o_3$.

\begin{figure}[t]
    \centering
    \includegraphics[width=\linewidth,trim=0 0 0 0,clip]{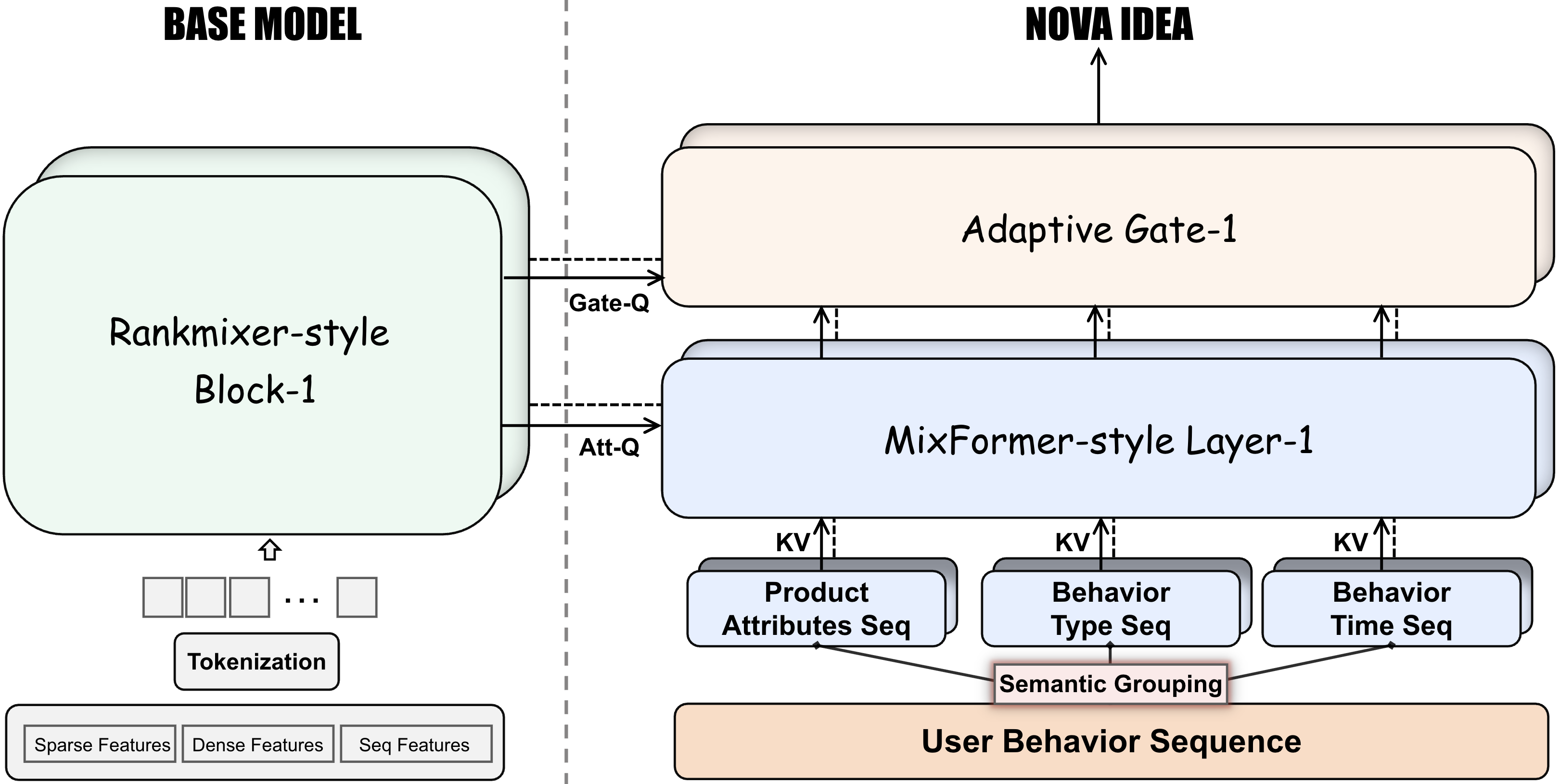}
    \caption{RankMixer (left) and NOVA's MixFormer adaptation (right). Semantic grouping yields three behavior sequences; each layer applies parallel cross-attention and gated residual fusion using queries from non-sequential RankMixer tokens.}
    \Description{The base model stacks RankMixer-style blocks over tokenized features. NOVA groups user-behavior side information into product-attribute, behavior-type, and behavior-time sequences. At each layer, the non-sequential RankMixer output queries the three sequences through parallel cross-attention and controls an adaptive gate that fuses the resulting contexts with a residual update.}
    \label{fig:mixformer_multiroute}
\end{figure}

\paragraph{Case 2: Multi-sequence MixFormer adaptation.}
Figure~\ref{fig:mixformer_multiroute} groups behavior side information into product-attribute, behavior-type, and behavior-time sequences. At each layer, non-sequential RankMixer outputs supply attention and gating queries; parallel MixFormer cross-attention models the three sequences, and adaptive gating fuses their outputs before a residual update. Relative to MixFormer, the design adds semantic subsequence decomposition and gated multi-sequence fusion.

The selected variant improves offline AUC by $0.05$, $0.02$, and $0.02$ pp on the top three objectives, respectively, with $3.5\%$ more parameters than RankMixer. Its gains remain below the $+0.15$ pp promotion threshold, so it was not advanced to online A/B testing.

\subsection{Online Validation for RQ4}
\label{subsec:online_ab}
The Round~6 architecture from Case~1 (Table~\ref{tab:case1_trajectory}) is selected for online validation. We evaluate the resulting pCVR model in a six-day, user-randomized A/B test on $5\%$ of production traffic. For pCVR objective $o$, we define the predicted-over-observed conversion ratio and its absolute bias as
\begin{equation}
\mathrm{PCOC}_o
=\frac{\sum_{i\in\mathcal{D}_o}\widehat{p}_{i,o}}
{\sum_{i\in\mathcal{D}_o}y_{i,o}},
\qquad
\mathrm{Bias}_o=\left|\mathrm{PCOC}_o-1\right|,
\end{equation}
where $\mathcal{D}_o$ is the evaluation set, $\widehat{p}_{i,o}$ is predicted pCVR, and $y_{i,o}$ is the observed conversion indicator, so $\sum_i y_{i,o}$ is the observed conversion count. Using $t$ and $c$ for treatment and control, the relative reduction in absolute bias is $1-\mathrm{Bias}_{o,t}/\mathrm{Bias}_{o,c}$. Table~\ref{tab:online_ab} reports treatment lifts relative to control.

\begin{table}[t]
\centering
\footnotesize
\setlength{\tabcolsep}{4pt}
\caption{Six-day user-randomized A/B results on $5\%$ of production traffic.}
\label{tab:online_ab}
\renewcommand{\arraystretch}{1.15}
\begin{tabular}{lcc}
\toprule
\textbf{pCVR objective} & \textbf{GMV lift} & \shortstack{\textbf{Relative bias}\\\textbf{reduction}} \\
\midrule
$o_1$ & $+1.25\%$ & $58.8\%$ \\
$o_2$ & $+1.70\%$ & $66.7\%$ \\
$o_3$ & $+2.02\%$ & $37.3\%$ \\
\bottomrule
\end{tabular}
\end{table}

GMV gains are positive and statistically significant across all three objectives, ranging from $1.25\%$ to $2.02\%$; relative bias reductions are $37.3\%$--$66.7\%$. Based on these results, we plan to proceed with full deployment of this architecture.

\subsection{Deployment Lessons and Limitations}
\label{subsec:deployment_lessons}

Using NOVA in production highlights two practical considerations: artifact reuse improves iteration efficiency, while request specificity affects modification quality.

\paragraph{Artifact reuse.}
NOVA reuses cached artifacts across tasks. Material Understanding outputs and production-backbone analyses are keyed by arXiv URL and backbone-code ID, respectively. For a repeated task using an already analyzed paper and backbone, retrieving both artifacts reduces the per-task LLM cost of the Init stage (Material~\&~Backbone Understanding) by $83.8\%$ (Appendix Table~\ref{tab:llm_usage}).

\paragraph{Request specificity.}
Modification quality remains sensitive to request specificity. For sequence-modeling tasks, the knowledge base contains multiple valid patterns for side-information pooling and sequence--non-sequence fusion (Appendix~\ref{app:knowledge-base}). An underspecified request may therefore produce a semantically valid implementation that differs from the engineer's intent; explicit architectural choices improve alignment.

%% file: 6Conclusion.tex
\section{Conclusion}
\label{sec:conclusion}
We present \textsc{NOVA}, a verification-aware agent harness for modifying production recommender architectures under deployment constraints. NOVA generates multiple candidates, rejects semantic violations before training, and uses trajectory memory to guide subsequent rounds. Across ScaleUp and Literature-to-Production tasks, NOVA achieves the highest effective pass rates among the evaluated baselines. Its offline-selected architecture also increases GMV and reduces absolute pCVR bias across three major objectives in a live A/B test.

%% file: 8Appendix.tex
\section{Appendix: Implementation, Audit, and Reproducibility Details}
\label{sec:appendix}

\newtcolorbox{promptbox}[1]{%
  enhanced, breakable,
  colback=white, colframe=black!55,
  colbacktitle=black!10, coltitle=black,
  fonttitle=\bfseries\small,
  attach boxed title to top left={xshift=4mm, yshift=-2.5mm},
  boxed title style={colframe=black!55, colback=black!10, sharp corners,
                     boxrule=0.3pt, size=small},
  title={#1},
  boxrule=0.4pt, arc=1pt,
  left=4pt, right=4pt, top=6pt, bottom=4pt,
  fontupper=\ttfamily\footnotesize,
  before upper={\setlength{\parindent}{0pt}\setlength{\parskip}{2pt}},
  enlarge top by=2mm,
}

This appendix has four parts. We report per-stage LLM usage and cost, expose the Quality Review mechanism through representative semantic rules and their audit trail, present reproducibility artifacts that reveal NOVA at the mechanism level without releasing proprietary production prompts verbatim, and describe the organization and incremental construction of its knowledge bases.

\subsection{Per-Stage LLM Usage and Cost}
\label{sub:harness_layout}
Table~\ref{tab:llm_usage} reports NOVA's average Claude API usage and cost by workflow stage.

Token usage and cost are dominated by Code Generation and the \textbf{Initialization} stage: the former produces the largest edits under the richest production context, while the latter ingests the source paper and deployed codebase. Local Testing and Offline Training are executed primarily by deterministic scripts; their small LLM usage is limited to interpreting compact status and error summaries. Offline metrics are computed deterministically; the LLM usage reported for Offline Evaluation reflects result interpretation and structured-feedback synthesis. The Main Agent only orchestrates and never calls an LLM directly.

\begin{table}[!htbp]
\centering\footnotesize
\setlength{\tabcolsep}{2.5pt}
\renewcommand{\arraystretch}{1.05}
\caption{NOVA per-stage Claude API usage. All figures are \emph{per-task averages} accumulated over all tool-call rounds within a task. \textbf{Initialization} denotes material and production-backbone understanding at task entry. \textbf{Uncached Input Tokens} and \textbf{Cache-Read Input Tokens} are the two Claude API input fields, with cache-read input billed at about $10\%$ of the base input rate. \textbf{Cost} is computed from the reported token usage and token prices.}
\label{tab:llm_usage}
\begin{tabularx}{\columnwidth}{@{}>{\raggedright\arraybackslash}X rrrr@{}}
\toprule
\multirow{2}{*}{Stage} & \multicolumn{3}{c}{Token usage} & \multirow{2}{*}{Cost} \\
\cmidrule(lr){2-4}
& Uncached & Cache read & Output & \\
\midrule
Initialization                                                                 & 797.7K & 14.45M &  88.7K & \$8.17 \\
Initialization + artifact reuse                                                & 137.2K &  2.24M &  14.7K & \$1.32 \\
Solution Design                                                                & 570.3K &  6.85M &  52.0K & \$4.68 \\
Code Generation                                    & 903.6K & 17.13M &  90.0K & \$9.30 \\
Quality Review                                     & 520.0K &  9.20M &  55.0K & \$4.80 \\
Local Testing                                      & 120.0K &  1.60M &   8.0K & \$1.05 \\
Offline Training                                   &  40.0K &  0.35M &   2.5K & \$0.35 \\
Offline Evaluation                                 &  98.0K &  1.34M &   6.3K & \$0.79 \\
\bottomrule
\end{tabularx}
\end{table}

\subsection{Representative Semantic Rules and Audit Trail}
\label{app:semantic-rules}
As introduced in Section~\ref{subsec:quality-review}, $V_{\mathrm{sem}}$ checks candidates against paper requirements, rules derived from semantic errors confirmed in earlier rounds, and validated rules from $KB$. Table~\ref{tab:semantic-rules-examples} gives one representative rule from each source. The first three columns instantiate the rule fields defined in Section~\ref{subsec:quality-review}; the final column shows the diagnostic emitted when a condition is violated. These examples illustrate the rule structure and are not used to estimate the reported semantic failure rate.

\begin{table*}[t]
\centering
\footnotesize
\setlength{\tabcolsep}{3pt}
\renewcommand{\arraystretch}{1.0}
\caption{Representative semantic rules used by $V_{\mathrm{sem}}$, drawn from the three sources described in Section~\ref{subsec:quality-review}. Rule identifiers are anonymized.}
\label{tab:semantic-rules-examples}
\begin{tabularx}{\textwidth}{p{0.14\textwidth} p{0.17\textwidth} >{\raggedright\arraybackslash}X p{0.19\textwidth}}
\toprule
Source & Scope & Condition (violation triggers \texttt{reject}) & Diagnostic emitted \\
\midrule
Paper requirement
& Feature-token mapping
& Features assigned to the same token must share semantic meaning; adding or moving a feature must preserve the paper-specified grouping.
& Rule ID and affected features or tokens \\

Earlier-round error
& Auxiliary-loss routing
& Each auxiliary loss must use the prediction and label indices of its design-specified objective (Table~\ref{tab:case1_trajectory}).
& Rule ID and affected loss, head, or index \\

$KB$ rule
& Sequence interaction
& When item-to-item interaction is required, a Transformer sequence block must not replace self-attention with a token-wise MLP that cannot exchange information across items.
& Rule ID and affected module or code location \\
\bottomrule
\end{tabularx}
\end{table*}

Paper requirements and $KB$ rules are loaded at task entry. Within the current task, rules derived from confirmed errors are applied in all subsequent rounds. The $KB$ snapshot remains fixed throughout the task. After task completion, rules validated as reusable may be incorporated into future $KB$ snapshots.

\paragraph{Audit trail and relation to $\mathrm{SFR}_{\mathrm{sem}}$.}
Each invocation of $V_{\mathrm{sem}}$ records its decision, applicable rule identifiers, and code locations. These automatic reports drive candidate filtering and are retained as trajectory evidence, but they are not used directly as the reported $\mathrm{SFR}_{\mathrm{sem}}$ labels. For comparable evaluation, three experts independently audit each candidate against the same rule sources, and majority vote determines the candidate's final semantic label. A round counts as a semantic failure only if every audited candidate in that round receives an invalid label. $N_{\mathrm{sem}}^-$ counts such rounds, and $\mathrm{SFR}_{\mathrm{sem}}=N_{\mathrm{sem}}^-/N_{\mathrm{iter}}$. Thus, the search-time reports make NOVA's automatic decisions auditable, while the expert labels provide a common evaluation criterion across methods.

\subsection{Reproducibility Artifacts}
\label{sub:reproducibility}
Because the production prompt corpus contains proprietary operational rules and identifiers, we do \emph{not} release it verbatim. Instead, we support reproducibility at the mechanism level through \textbf{four artifact types}: (i) a user task template (\S\ref{sub:user_task}) showing the inputs parsed before Solution Design; (ii) a prompt template skeleton (\S\ref{sub:llm_prompt}) exposing how an agent consumes structured guidance $z_t$; (iii) a semantic-rule update and trajectory snippet (\S\ref{sub:semantic_traj}) showing how a newly confirmed error affects subsequent verification; and (iv) representative skill summaries (\S\ref{sub:skill_examples}) illustrating stage-specific procedural decomposition. Together, these artifacts reveal NOVA's core mechanisms without disclosing proprietary content.

\subsubsection{User Task Template}
\label{sub:user_task}
Every NOVA task starts from a structured user request that fixes the base model, the target metric with a threshold, the resource budget, and the data window. This template lets the Main Agent route the task to the correct capability level before Solution Design (Section~\ref{subsec:implementation}), without any follow-up dialog. We show a compressed L3 example below; identifiers and paths are anonymized. Here $\tau$ (early-stop / primary-success threshold) and $R_{\max}$ (max rounds per task) match the main-text configuration, and $\varepsilon$ is the task-level guard margin (the maximum allowed drop on any other metric).

\begin{promptbox}{User Task Template (L3, Literature-to-Production request)}
\# TASK\\
\hphantom{xx}base\hphantom{xxx}: <topo\_id> \hspace{1em} resources: <group\_ids>\\[3pt]
\# REQUIREMENT\\
\hphantom{xx}paper\hphantom{x}: <arXiv link>\\
\hphantom{xx}goal\hphantom{xx}: adapt task-token decoupling (TIM) onto\\
\hphantom{xxxxxxxxx}the deployed RankMixer backbone.\\
\hphantom{xx}window : <data\_window>\\[3pt]
\# METRICS\\
\hphantom{xx}primary: $\Delta\mathrm{AUC}_{\mathrm{pp}}^{(0)}\ge \tau$ on $\ge 2$ of\\
\hphantom{xxxxxxxxx}$\{$og\_a,...,og\_d$\}$, including og\_a\\
\hphantom{xx}guard\hphantom{xx}: no other og drops by more than $\varepsilon$\\[3pt]
\# SEARCH BUDGET\\
\hphantom{xx}$R_{\max}$ (max rounds per task)\\[3pt]
\# CONSTRAINTS\\
\hphantom{xx}implementation / resource / workflow constraints ...
\end{promptbox}

\subsubsection{Prompt Template Skeleton}
\label{sub:llm_prompt}
We reproduce below a demonstrative skeleton of the NOVA Solution Design prompt, with proprietary identifiers and task-specific content redacted. The proposal-generation prompt and the $V_{\mathrm{sem}}$ reviewer prompt share this same template, differing only in the role/task slot. It follows the interface in Section~\ref{subsec:search}: the Main Agent supplies the selected capability level and structured inputs, and Solution Design returns $K$ modification plans for Code Generation. Semantic verification and candidate ranking use separate Quality Review prompts.

\begin{promptbox}{Solution Design Agent Prompt}
\# ROLE\\
\hphantom{xx}The Solution Design agent in NOVA: each round it consumes\\
\hphantom{xx}the capability level $\ell\in\{$L1,...,L4$\}$ selected by the Main\\
\hphantom{xx}Agent and the outputs of prior stages, invokes the matching\\
\hphantom{xx}skill, and returns $K$ ranked modification plans for Code Generation.\\[3pt]
\# INPUTS\\
\hphantom{xx}$e_{\mathrm{prev}}$: previous modification plan\\
\hphantom{xx}$V = \{$sem, loc, failure\_loc$\}$:\\
\hphantom{xxxx}semantic / local / failure-localisation reports\\
\hphantom{xx}$\Delta J$: metric-gain set\\
\hphantom{xxxx}$\{\Delta\mathrm{AUC},\Delta\mathrm{DelayAUC},\Delta\mathrm{Params},\Delta\mathrm{FLOPs}\}$\\
\hphantom{xx}$H$: trajectory memory (SUMMARY + prior EVALUATION)\\[3pt]
\# OUTPUT\\
\hphantom{xx}header $z = \{$weak\_components, directions, forbidden$\}$\\
\hphantom{xx}+ DESIGN/design.md with $K$ candidate plans\\
\hphantom{xx}(target module, expected $\Delta J$, guardrails, runnable diff);\\
\hphantom{xx}each plan must satisfy $\Omega$ and avoid any forbidden direction.\\[3pt]
\# ITERATION\\
\hphantom{xx}for $t>0$, read prior EVALUATION and the cumulative SUMMARY,\\
\hphantom{xx}and refine directions rather than regenerating from scratch.
\end{promptbox}

\subsubsection{A Semantic-Rule Update and a Trajectory Snippet}
\label{sub:semantic_traj}
Panel (a) shows the semantic rule confirmed in Case~1, and panel (b) shows how the corresponding trajectory evidence makes the rule available to subsequent rounds. This rule belongs to the active semantic-rule set $\mathcal L_t$; it is distinct from the task-scoped ineffective patterns in $z_t^{\mathrm{forbid}}$.

\begin{promptbox}{(a) Semantic Rule Confirmed in Round 4}
\{\\
\hphantom{xx}rule\_id\hphantom{xxxx}: "auxiliary-loss-objective-routing",\\
\hphantom{xx}source\hphantom{xxxxx}: "confirmed earlier-round error",\\
\hphantom{xx}scope\hphantom{xxxxxx}: "auxiliary-loss routing",\\
\hphantom{xx}condition\hphantom{xx}: "prediction and label indices must\\
\hphantom{xxxxxxxxxxxxxxxx}match the design-specified objective"\\
\}
\end{promptbox}

\begin{promptbox}{(b) Excerpted Trajectory (rounds 3--5)}
\# t = 3\\
\hphantom{xx}$c_{3,j_3^*}$ passes the active rules and local tests.\\
\hphantom{xx}$\to$ Offline results suggest that the $o_1$-specific loss\\
\hphantom{xxxxx}may use mismatched prediction/label indices.\\[3pt]
\# t = 4\\
\hphantom{xx}Code inspection confirms the objective-indexing error.\\
\hphantom{xx}$\to$ The implementation is corrected and the semantic rule\\
\hphantom{xxxxx}in (a) is derived from the confirmed error.\\
\hphantom{xx}$\to$ The candidate report and outcome are retained in\\
\hphantom{xxxxx}$\mathcal O_5$.\\[3pt]
\# t $\geq$ 5\\
\hphantom{xx}$\mathcal L_t$ includes the new auxiliary-loss routing rule.\\
\hphantom{xx}$\to$ $V_{\mathrm{sem}}(c_{t,j};\mathcal L_{t,j})$ rejects later\\
\hphantom{xxxxx}candidates that repeat the confirmed semantic error.
\end{promptbox}

\subsubsection{Skill Library Examples}
\label{sub:skill_examples}
NOVA decomposes architecture evolution into stage-specific agents that invoke reusable, file-grounded skills. Below we show compressed summaries of two representative skills. Each summary keeps the operational structure --- inputs, core workflow, representative guardrails, and outputs --- while omitting long rule lists and proprietary field names.

\begin{promptbox}{Material Understanding Skill (Initialization Stage)}
\# INPUT\\
\hphantom{xx}Task request, paper PDF, production backbone, target framework.\\[2pt]
\# PROCESS\\
\hphantom{xx}1. Extract architecture, equations, shapes, module dependencies.\\
\hphantom{xx}2. Tag each item as paper-stated / inferred / engineering assumption;\\
\hphantom{xxxxx}align paper modules with backbone integration points.\\
\hphantom{xx}3. Emit a structured spec and log unresolved ambiguities.\\[2pt]
\# OUTPUT\\
\hphantom{xx}\{material\_brief, architecture\_map, equation\_map,\\
\hphantom{xxxx}shape\_constraints, paper\_code\_alignment, ambiguity\_log\}.md
\end{promptbox}

\begin{promptbox}{Architecture Planning Skill (Solution Design)}
\# INPUT\\
\hphantom{xx}$q,A_t,z_t,KB,\Omega,\ell,K$ (+ material artifacts).\\[2pt]
\# PROCESS\\
\hphantom{xx}1. Ground $q$ in $A_t$.\\
\hphantom{xx}2. Generate $K$ distinct plans using $z_t$ and $KB$.\\
\hphantom{xx}3. Attach components, steps, effects, assumptions, and\\
\hphantom{xxxxx}checks; exclude $z_t^{\mathrm{forbid}}$ patterns.\\[2pt]
\# OUTPUT\\
\hphantom{xx}$(e_{t,1},\ldots,e_{t,K})$; implementation and ranking remain\\
\hphantom{xx}downstream.
\end{promptbox}

\setcounter{dbltopnumber}{2}
\begin{table*}[t]
\centering
\footnotesize
\setlength{\tabcolsep}{5pt}
\renewcommand{\arraystretch}{1.08}
\caption{Representative P02 variants for sequence side-information pooling. $B$, $L$, $D$, $N_f$, and $M$ denote batch size, sequence length, input width, the number of side-information fields, and the number of retained temporal segments, respectively. In P02d, $N$ input shards are reduced into $G$ semantic groups and projected to width $d$.}
\label{tab:seq-p02-variants}
\begin{tabularx}{\textwidth}{@{}>{\raggedright\arraybackslash}p{0.22\textwidth} >{\raggedright\arraybackslash}X >{\raggedright\arraybackslash}p{0.27\textwidth}@{}}
\toprule
Variant & Aggregation operation & Output shape \\
\midrule
P02a: Within-step reduction
& Aggregate side-information fields independently at each time step.
& Preserves sequence length: $[B,L,N_f{\times}D]\rightarrow[B,L,D]$. \\
P02b: Masked temporal pooling
& Apply masked sum or mean pooling over the temporal axis.
& Produces one sequence vector: $[B,L,D]\rightarrow[B,D]$. \\
P02c: Segmented window pooling
& Pool nonuniform temporal windows separately.
& Retains coarse temporal order as $M$ tokens: $[B,L,D]\rightarrow[B,M,D]$. \\
P02d: Semantic-group reduction
& Aggregate aligned side-information fields within semantic groups and project the grouped representations.
& Preserves sequence length across $G$ groups: $\{[B,L,D]\}_{n=1}^{N}\!\rightarrow[B,L,GD]\!\rightarrow[B,L,d]$. \\
\bottomrule
\end{tabularx}
\end{table*}

\begin{table*}[t]
\centering
\footnotesize
\setlength{\tabcolsep}{5pt}
\renewcommand{\arraystretch}{1.08}
\caption{Representative P10 variants for MixFormer-style fusion between sequence representations and non-sequential backbone tokens.}
\label{tab:seq-p10-variants}
\begin{tabularx}{\textwidth}{@{}>{\raggedright\arraybackslash}p{0.22\textwidth} >{\raggedright\arraybackslash}X >{\raggedright\arraybackslash}p{0.27\textwidth}@{}}
\toprule
Variant & Interaction design & Key property \\
\midrule
Heavy MixFormer
& Use non-sequential tokens as queries and variable-length sequence representations as keys and values, with FlashAttention and grouped-query attention.
& High-capacity interaction with the greatest engineering complexity. \\
Lightweight UniFormer
& Replace the heavy attention path with manual attention over padded sequences.
& Simpler single-sequence injection. \\
Dual-sequence gated XDUniformer
& Model multiple sequence types through parallel attention paths and combine their contexts by gating.
& Supports heterogeneous sequences with adaptive fusion. \\
Dual-stage MixFormerV2
& Refine sequence representations before cross-attention with backbone tokens.
& Adds sequence-side interaction before backbone injection. \\
Dense MixFormer
& Combine manual attention with lightweight normalization, dense feed-forward blocks, and output fusion.
& Compact static-sequence implementation with moderate engineering cost. \\
\bottomrule
\end{tabularx}
\end{table*}

\subsection{Knowledge-Base Organization and Incremental Construction}
\label{app:knowledge-base}

NOVA maintains task-specific knowledge bases for feature-editing specifications, sequence modeling, basic-feature modeling, and other architecture-modification tasks. The Main Agent retrieves the relevant stores to construct the fixed snapshot $KB$ used during search.

\paragraph{Content--construction separation.}
Each knowledge base separates queryable \texttt{db\_content} from its construction workflow and evidence in \texttt{db\_construction}. For new code, a construction skill traces configuration and execution, aligns feature declarations with their uses, and maps the implementation to the existing taxonomy. Only a genuinely new modeling pattern or validated variant is promoted to \texttt{db\_content}; otherwise, the existing entry is reused and the analysis remains construction evidence. This keeps retrieval concise and future updates auditable.

\paragraph{Sequence-modeling example.}
The sequence-modeling knowledge base links feature configurations to modeling patterns, integration paradigms, and validated code cases. It currently contains six sequence-configuration patterns and ten modeling patterns. Tables~\ref{tab:seq-p02-variants} and~\ref{tab:seq-p10-variants} illustrate its granularity through P02, which captures alternative ways to pool sequence side information, and P10, which captures MixFormer-style interactions between sequence representations and non-sequential backbone tokens.